\documentclass[preprint2 ]{aastex}

\usepackage{graphicx,color}
\usepackage{hyperref}
\usepackage{xspace}
\usepackage{longtable}
\usepackage{amsmath}

\newcommand{\kms}{km s$^{-1}$\xspace}
\newcommand{\HI}{{\rm H\,{\scriptsize I}}\xspace}
\newcommand{\HII}{{\rm H\,{\scriptsize II}}\xspace}
\newcommand{\noprint}[1]{}

\shorttitle{\HI Absorption towards \HII Regions in the SGPS}
\shortauthors{Brown, C., et al.}

\begin{document}
\title{A Complete Atlas of \HI Absorption toward \HII Regions \\in the Southern Galactic Plane Survey (SGPS I)}
\author{C. Brown\altaffilmark{1,2} , J. M. Dickey\altaffilmark{1}, J. R. Dawson\altaffilmark{2,1} \& N. M. McClure-Griffiths\altaffilmark{2}}
\affil{1. School of Physical Sciences, Private Bag 37, University of Tasmania, Hobart, 7001, Australia}
\affil{2. CSIRO Astronomy and Space Science, ATNF, PO Box 76, Epping, NSW, 1710, Australia}

\begin{abstract}
We present a complete catalog of \HI emission and absorption spectrum pairs, toward \HII regions, detectable within the boundaries of the Southern Galactic Plane Survey (SGPS I), a total of 252 regions.  The catalog is presented in graphical, numerical and summary formats.  We demonstrate an application of this new dataset through an investigation of the locus of the Near 3kpc Arm.

\end{abstract}

\keywords{galaxy: HII regions}

\section{Introduction}
Combining data from the Australia Telescope Compact Array (ATCA) and the Parkes single dish telescopes, the SGPS  \citep[SGPS,][]{SGPS} provides \HI line and 1.4 GHz radio continuum data for the fourth Galactic quadrant, with the best combination of resolution and sensitivity currently available for that line in this area.  Using these data, we have measured 21-cm absorption spectra toward a sample of 252 \HII regions.  

This study produces two distinct data products: a consolidated census of \HII regions with known radio recombination line (RRL) velocities, taken from the literature; as well as the \HI emission and absorption spectrum pairs towards them - within the bounds of the SGPS (255\arcdeg$<l<$353\arcdeg).  These two resources will enable further study into the structure and dynamics of the neutral interstellar medium (ISM) in the fourth quadrant (on which massive stars and their formation have a significant impact).  The catalogs will serve as a data set for numerous studies; including investigations of the spiral structure of the Galaxy \citep[eg. ][]{Strasser07} and kinematic distance works, both for target regions \citep[eg.][]{Urquhart12} as well as for the intervening \HI clouds \citep[eg.][]{RomanDuval09}.

The sample selection from existing \HII region catalogs is described in Section \ref{Data}, while the method of spectrum extraction appears in Section \ref{extract}.  The catalog itself appears in Section \ref{Catalog}, before a disucssion of the global properties of \HI absorption in Section \ref{Discussion} and an illustration of its use: an examination of the distribution of \HI absorption in longitude-velocity ($lv$) space, including the locus of the Near 3kpc Arm.

\section{Data}
\label{Data}

\subsection{Southern Galactic Plane Survey, SGPS \label{SGPSdata}}
The Southern Galactic Plane Survey covers 325 square degrees of the Galactic plane over the fourth and first Galactic quadrants (SGPS I and SGPS II respectively).  For this work, SGPS refers to the fourth quadrant component of the survey only (i.e. the SGPS I).  The SGPS I provides both the \HI line data, as well as the continuum maps \citep{Haverkorn06} used for source detection and identification. 

The SGPS provides three distinct data products: Parkes continuum-subtracted cubes, with an angular resolution of $15\arcmin$; combined Parkes and Australia Telescope Compact Array (ATCA) continuum-subtracted \HI cubes (2.\arcmin2) and combined Parkes and ATCA cubes containing continuum emission ($\sim1.\arcmin6$).  For this work we use the continuum \textit{included} combined Parkes-ATCA data.  These cubes were specifically produced for \HI absorption studies, as they provide accurately calibrated data at the highest angular resolution of the SGPS.  All SGPS continuum and line cube data is available online via the Australia Telescope National Facility (ATNF) \HI Surveys archive \footnote{\url{http://www.atnf.csiro.au/research/HI/common}}.

SGPS data has been used extensively in a number of fields, including \HI self-absorption \citep{Gibson05, Kavars05}, the cold neutral interstellar medium in the outer Galaxy \citep{Strasser07}, Galactic structure \citep{McCGDistantArm} as well as investigations of particular individual sources \citep{KothesDougherty07}.

\subsection{\HII Regions \label{HIIregions}}
Firstly we compiled a list of \HII regions from RRL catalogs for the longitude and latitude range of the SGPS. This compilation of \HII regions was then over-plotted on the SGPS radio continuum maps using the KARMA software suite \citep{Gooch95}.  Each map was then visually inspected to confirm \HII region detection.  Target \HII regions were included in this catalog (\S \ref{Catalog})  if a \textit{single, distinct} emission source was visible in the SGPS continuum map.

Several cataloged \HII regions were coincident with several radio continuum sources and some targets had coordinates coincident with no continuum source.   It is not surprising that the SGPS with angular resolution of $\sim2\arcmin$ detects multiple sources within the larger beam of the early \HII region discovery works, for instance the Parkes beam of \citet{CH87} is $\sim4\arcmin$.  Furthermore, there are often clusters of continuum emission sources surrounding a cataloged region\rq{}s coordinates, such that no attempt can be made to distinguish the background emission (see \S \ref{extract}).  There are several \HII regions that are included in more than one \HII region RRL velocity catalog (\S \ref{HIIcatalogs}) and therefore become duplicate sources when the catalogs are combined.  Furthermore, there were \HII regions that did not appear as continuum sources in the SGPS, or had continuum temperatures $T_{\rm{cont}}<5$K, these were not included in this catalog. 
 
The total number of individual, distinct \HII regions, visible in the SGPS continuum is 252.  Throughout this work, the name (in form ``G longitude $\pm$latitude'') reported for each \HII region is taken from the catalog from which the region is provided (see \S \ref{HIIcatalogs}); hence the inconsistencies in decimal precision.

\subsubsection{\HII Region Catalogs \label{HIIcatalogs}}
We use the catalog of \citet{CH87} as the basis of our target list, supplementing that catalog with further \HII regions sourced from other works.  Our final source list is a compilation of the \HII region RRL velocity catalogs of \citet{CH87}, the Green Bank Telescope \HII Region Discovery Survey \citep{GBTHRDS}, \citet{L89} and \citet{Wilson70} (see Table \ref{Tablecatalog}).  We describe each of these works below.  

There are several other fourth quadrant studies which include \HII region candidates \citep{Reifenstein70,Walsh99,Busfield06,Urquhart12, MALT90}, but as Hn$\alpha$ emission has not been detected towards these candidates, they are not confirmed \HII regions and are therefore excluded from this work.

\paragraph{\citet{CH87}}
Hydrogen RRL parameters for 316 \HII regions, observed with the Parkes 64-m Radio Telescope, over the longitude range $210\arcdeg<l<360\arcdeg$ are provided by \citet{CH87}.  The majority of \HII regions in this study were sourced from this work, the southern correspondent to the \citet{L89} catalog.  Several of the \HII regions cataloged by \citet{CH87} are outside the bounds of the SGPS; therefore for regions with Galactic longitudes $l<245\arcdeg$, or $l>353\arcdeg$, or Galactic latitudes $|b|\gtrsim 1.4\arcdeg$ we cannot extract \HI E/A spectra (and they are not included in this catalog).  

\paragraph{Green Bank Telescope \HII Region Discovery Survey}
The Green Bank Telescope \HII Region Discovery Survey \citep[GBTHRDS,][]{GBTHRDS}\footnote{\url{http://www.cv.nrao.edu/hrds/}} detected 602 RRL components towards 448 continuum sources in the Galactic plane between $344\arcdeg<l<67\arcdeg$.  The survey more than doubled the number of known \HII regions within that longitude range  - the majority of previously known \HII regions were sourced from the \citet{L89} catalog, see below.  With a 95\% detection rate the GBTHRDS selected its target sample from spatially coincident mid-infrared and radio continuum emission. The \textit{Spitzer} MIPSGAL survey \citep{MIPSGAL} provided 24$\mu$m data, while 21cm continuum emission was sourced from the SGPS, VGPS \citep{VGPS} and NVSS \citep{NVSS}.  

\paragraph{\citet{L89}}
The canonical northern hemisphere catalog of \citet{L89} extends into the fourth quadrant, and therefore into the SGPS coverage area.  This large survey provides RRL velocity detections for 462 \HII regions.

\paragraph{\citet{Wilson70}}
\citet{Wilson70} detected the H$109\alpha$ RRL toward 130 \HII regions visible in the southern sky ($261\arcdeg<l<50\arcdeg$) using the Parkes Radio Telescope in 1968.

\section{Emission/Absorption Method of \HI Absorption Spectrum Extraction }
\label{extract}
\begin{figure}[h]
\centering
\includegraphics[width=\columnwidth]{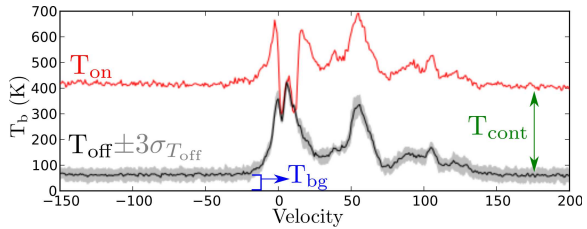}
\caption{Brightness temperature of sample $T_{\rm{on}}$ (peak brightness, shown in red) and average off source (black) spectra for G269.133-1.137.  The grey envelope around the average off source spectrum represents a $3\sigma$ variation between the three off source positions.  Components of Equations \ref{ontemp} and \ref{offtemp} are labeled. (A color version of this figure is available in the online journal.)}
\label{Fig1}
\end{figure}

The Emission/Absorption (E/A) method observes \HI spectra coincident with, and adjacent to, discrete continuum sources.  In order to derive absorption, $e^{-\tau}$, the brightness temperature as a function of velocity, $V$, both on ($T_{\rm{on}}$) and off the source (i.e. the emission spectrum, $T_{\rm{off}}$) are compared.  The simplest radiative transfer situation gives:
\begin{eqnarray}
T_{\rm{on}}(V)=(T_{\rm{bg}}+T_{\rm{cont}})e^{-\tau(V)}+T_{\rm{s}}(V)(1-e^{-\tau(V)}) \label{ontemp}\\
T_{\rm{off}}(V)=T_{\rm{bg}}e^{-\tau(V)}+T_{\rm{s}}(V)(1-e^{-\tau(V)}) \label{offtemp}
\end{eqnarray}  

where $T_{\rm{cont}}$ is the continuum source brightness temperature, $T_{\rm{s}}$ is the spin temperature of the foreground cloud(s) and $T_{\rm{bg}}$ represents the brightness temperature of any other background contribution (see Figure \ref{Fig1}).

Assuming that the `on' and `off' spectra both sample the same gas, subtraction of one from the other removes the common $T_{\rm{bg}}$ and $T_{\rm{s}}(V)$ terms, allowing the absorption to be calculated directly,

\begin{equation}
e^{-\tau}=\frac{T_{\rm{on}}-T_{\rm{off}}}{T_{\rm{cont}}} \label{EqAbs}
\end{equation}

The `on' and three `off' source positions, which were averaged to provide an emission estimate, were chosen in accordance with the criteria identified in \citet{JonesDickey12}.  Large scale \HI emission fluctuations, which are present on all angular scales \citep{Green93, Dickey01}, are reflected by the variation in the three `off' source positions, see \S \ref{Quals}.  The E/A method is described in detail by \citet{Kolpak03}.

\subsection{Continuum Calibration}\label{contincalc}
The continuum temperature is estimated from the difference in on and off spectra across a range of velocity channels in which there is no \HI signal.  For most spectra we used the velocity channel range 100\kms$<V<175$\kms as this avoids both expected Galactic circular rotation velocities as well as avoiding the ends of the spectrum band.  

The SGPS line data is recorded in units Jy/beam, therefore a conversion to K is required.  Firstly a two-dimensional Gaussian beam solid angle is assumed (Equation \ref{beam}), then the flux density for an unresolved source is approximated (Equation \ref{fluxdensity}) in terms of antenna temperature.

\begin{equation}
\Omega=\frac{\pi}{4 \ln(2)}(\rm{FWHM}_{\phi} \times \rm{FWHM}_{\psi})\label{beam}
\end{equation}
\begin{equation}
S=\frac{2k}{\lambda^2}\Omega T_A\label{fluxdensity}
\end{equation}

Theoretically, the antenna temperature is the convolution of the beam response with the sky brightness temperature distribution, integrated over the entire sky. Substituting for beam solid angle, $\lambda=21.1$cm and creating dimensionless variables results in Equation \ref{conversion}.

\begin{equation}
T_b\approx 606 \frac{S/\rm{(Jy/beam)}}{(\rm{FWHM}_\psi \times \rm{FWHM}_\phi)/arcseconds^2} \qquad \rm{K}\label{conversion}
\end{equation}

The dimensions of each synthesized beam are given in Table 4 of \citet{SGPS}.

This conversion factor is used to convert the brightness temperature of the off source spectrum (top panel) of each image in Figure \ref{Fig4} as well as to obtain the continuum temperature reported in Table \ref{Tablecatalog}.  However, the conversion is $not$ required in the calculation of absorption (see Equation \ref{EqAbs}).

\subsection{Quality of Spectra}\label{Quals}
As for most emission/absorption studies, the noise level in the absorption spectrum is not dictated by radiometer noise, but rather the precision with which the absorption spectrum can be subtracted from background continuum emission, see Equation \ref{EqAbs} \citep{Dickey03}.

\begin{figure}[h!]
\centering
\includegraphics[width=\columnwidth]{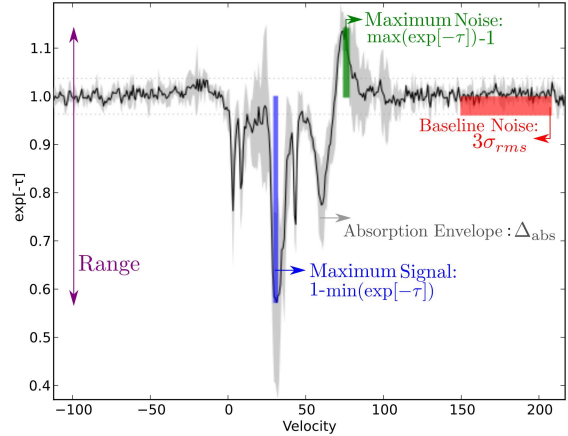}
\caption{Sample absorption spectrum for G254.676+0.229, $e^{-\tau}$ is the solid black line, demonstrating the quality tests.  The range (puprle), maximum signal (blue) and maximum noise (green), baseline noise (red) and absorption fluctuation envelope (grey) are shown. (A color version  of this figure is available in the online journal.)
\label{Fig2}}
\end{figure}

A series of five tests were devised to measure the quality of each \HI absorption spectrum  (see Figure \ref{Fig2})---resulting in six quality categories A-F. The quality category for each spectrum included in the catalog is given in Table \ref{Tablecatalog}.  Each spectrum was initially assumed to be at the best quality rating (i.e. category A); the quality factor was then down-graded for each test failed; therefore the spectra in category F failed all five quality tests listed below. 

\paragraph{Range}
Tests if the range of calculated absorption values are realistic: theoretically $0<e^{-\tau}<1$.  Continuum temperature uncertainties or uncertainties in assumed background emission spectrum will increase the range of $e^{-\tau}$.  This test is failed if Range($\exp[-\tau]$)$>1.5$.

\paragraph{Maximum Signal to Maximum Noise}
`Maximum signal' refers to $1-min(\exp[-\tau])$ and `maximum noise' refers to $max(\exp[-\tau])-1$, see Figure \ref{Fig2}.  This test is designed to ascertain if emission signals (which should have been removed) overwhelm absorption in the spectrum.  This test is failed if (maximum signal/maximum noise)$<3$.

\paragraph{Absorption Uncertainty Envelope}
This test investigates the fluctuations in the absorption envelope (caused by differences between the three emission, off source, spectra).    The absorption uncertainty envelope is calculated as follows:
\begin{equation}
\Delta_{\textrm{abs}}(v)=\left|e^{-\tau}-\left(\frac{T_{\rm{on}}-(T_{\rm{off}}+3\sigma_{T_{\rm{off}}})}{T_{\rm{cont}}}\right)\right|
\end{equation}
Where $\sigma_{T_{\textrm{off}}}$ is the standard deviation between the three emission spectra, for each velocity channel.  This test is failed if the standard deviation of this envelope, over the whole velocity range, is large: i.e. $3\sigma(\Delta_{\rm{abs}})>1$.  The absorption uncertainty envelope is shown as grey shading in each \HI absorption spectrum panel (see Figures \ref{Fig2} and \ref{Fig4}), while $\pm3\sigma_{T_{\rm{off}}}$ is shown as the grey shading in each \HI emission spectrum panel (Figure \ref{Fig4}).

\paragraph{Baseline Noise}
This test identifies absorption spectra with high levels of baseline noise.  This is achieved by investigating the standard deviation of absorption ($\sigma_{\rm{rms}}$) over the same velocity channel range from which the continuum temperature is determined, i.e. there is no \HI emission nor absorption signals (see \S \ref{contincalc}). This test is failed if $3\sigma_{\rm{rms}}>1$.

\paragraph{Number of Velocity Channels with Significant Absorption}
The final test is a count of the number of velocity channels which demonstrate statistically significant amounts of absorption.  That is,  the absorption is deeper than a combination of the baseline noise and the absorption envelope uncertainty: $e^{-\tau}(v)<(1-3\sigma_{\rm{rms}}-\sigma(\Delta_{\rm{abs}}))$ .  At least 15 channels of statistically significant absorption are required to pass this test.  Note that $\sigma(\Delta_{\rm{abs}})\propto3\sigma_{T_{\rm{off}}}$ (see Absorption Uncertainty Envelope test).\\

\begin{figure}[h!]
\centering
\includegraphics[width=\columnwidth]{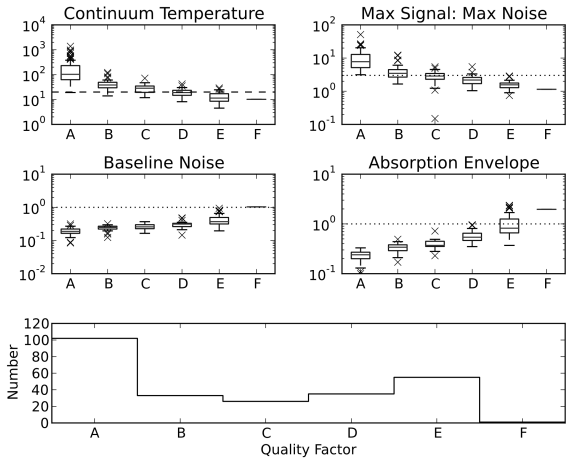}
\caption{Assessment of the quality control variables for each \HI absorption spectrum in categories A (best quality) to F (poorest quality). In each panel, the horizontal dotted line signifies the value for which the test is failed (see text).  The distribution of continuum temperatures in each quality category is also shown, even though contintuum temperatures do not constitute a formal quality test (\S \ref{Quals}); in this case the dashed horizontal line demonstrates the limit in continuum temperature required for resolution of the kinematic distance ambiguity from \citet{JonesDickey12}.  The bottom panel displays a histogram of the number of \HII regions in each quality category. (A color version  of this figure is available in the online journal.) \label{Fig3}}
\end{figure}

The majority of the poorest quality spectra (category F) come from target regions with $l>340\arcdeg$ and were usually sourced from the \HII region catalog of the GBTHRDS.  This is not particularly surprising as the GBTHRDS avoided previously-observed \HII regions; resulting in  intrinsically fainter sources compared to \citet{CH87}.

The bottom panel of Figure \ref{Fig3} displays the number of regions in each quality category.  The distribution appears to be superficially bimodal, such that the quality categories could be regrouped into `good' and `poor' umbrella categories associated with quality factors A-C and D-F respectively.  Through the use of these umbrella categories it can be shown that in order to extract a `good' quality \HI absorption spectrum, a continuum temperature greater than $\sim20$K is required---note:  while continuum temperature was not used as a quality assurance paramater, stronger sources are associated with higher quality factors (see Figure \ref{Fig3}).

\citet{JonesDickey12} reported that a continuum temperature of at least 10K is required to solve the kinematic distance ambiguity for \HII regions in the SGPS, using their velocity channel summation technique.  This limit approximately corresponds to the median continuum temperature of quality category E (10.4 K).  Note that \HII regions with continuum temperatures $T_{\rm{cont}}<5$ K were not included in the catalog of \HI spectra (\S \ref{HIIregions}).

\section{Catalog} \label{Catalog}
The catalog of \HI emission/absorption spectrum pairs is the primary data product of this work.  We present it here in three formats: individual spectrum figures, complete data set (in tar format) and summary table.

\paragraph{Catalog Summary}
Table \ref{Tablecatalog} provides an overview of the entire catalog.  The 252 included \HII regions are listed, along with their known parameters.   The columns list the Galactic coordinates of the \HII region (as reported by the original reference); SIMBAD identification; $T_{\rm{cont}}$ (K), quality factor, RRL velocity and  RRL velocity reference respectively.

\clearpage
\begin{deluxetable}{|lllcccc|}
\tablecolumns{7}
\tablewidth{0pt}
\tablecaption{\HII Region Catalog \label{Tablecatalog}}
\tabletypesize{\footnotesize}
\tablehead{
\multicolumn{1}{|l}{Longitude}&
\multicolumn{1}{l}{Latitude}&
\multicolumn{1}{l}{SIMBAD ID}&
\multicolumn{1}{l}{T$_{cont}$}&
\multicolumn{1}{c}{Qual}&
\multicolumn{1}{l}{V$_{RRL}$}&
\multicolumn{1}{c|}{Ref}}

\startdata
254.676&$+$0.299& \object{NGC 2579 }&A&134&64&1\\
263.619&$-$0.533& \object{GAL 263.62-00.53 }&A&58&1&1\\
267.947&$-$1.066& \object{GAL 267.95-01.07 }&A&969&1&1\\
268.0&$-$1.1& \object{GAL 268.0-01.1}&B&31&1.8&4\\
268.454&$-$0.848& \object{GAL 268.45-00.85 }&A&94&5&1\\
269.133&$-$1.137& \object{GAL 269.13-01.14 }&A&118&15&1\\
270.294&$+$0.848& \object{GAL 270.29+00.85 }&E&26&1&1\\
274.013&$-$1.141& \object{GUM 26 }&A&317&39&1\\
281.595&$-$0.969& \object{GAL 281.60-00.97 }&C&30&2&1\\
282.026&$-$1.181& \object{GAL 282.03-01.18 }&A&235&19&1\\
282.632&$-$0.853& \object{GAL 282.63-00.85 }&E&12&0&1\\
283.131&$-$0.984& \object{GAL 283.13-00.98 }&D&35&$-$1&1\\
283.312&$-$0.566& \object{GAL 283.31-00.57 }&E&20&6&1\\
283.329&$-$1.050& \object{GAL 283.33-01.05 }&C&26&16&1\\
284.308&$-$0.334& \object{GUM 29 }&A&879&0&1\\
284.559&$-$0.183& \object{GAL 284.56-00.18 }&B&29&9&1\\
284.650&$-$0.484& \object{GAL 284.65-00.48 }&E&17&5&1\\
284.723&$+$0.313& \object{GAL 284.72+00.31 }&A&72&10&1\\
285.253&$-$0.053& \object{GAL 285.25-00.05 }&A&321&$-$2&1\\
286.195&$-$0.163& \object{GAL 286.20-00.16 }&A&52&$-$18&1\\
286.873&$-$0.883& \object{GAL 286.87-00.88 }&D&24&$-$20&1\\
287.217&$+$0.053& \object{GAL 287.22+00.05 }&E&29&$-$25&1\\
287.247&$+$0.355& \object{GAL 287.25+00.36 }&E&22&$-$18&1\\
287.393&$-$0.630& \object{GAL 287.39-00.63 }&A&415&$-$17&1\\
287.550&$-$0.616& \object{GAL 287.55-00.62 }&A&344&$-$39, -14&1\\
289.063&$-$0.355& \object{GAL 289.06-00.36 }&A&88&19&1\\
289.755&$-$1.152& \object{GUM 35 }&B&38&22&1\\
289.878&$-$0.792& \object{2MASX J11005954-6050229 }&A&50&22&1\\
290.646&$+$0.256& \object{GAL 290.65+00.26 }&D&10&$-$28&1\\
291.059&$-$0.770& \object{GAL 291.06-00.77 }&B&41&17&1\\
291.284&$-$0.713& \object{OH 291.3 -0.7 }&A&1377&$-$25&1\\
291.466&$-$0.128& \object{GAL 291.47-00.13 }&E&22&6&1\\
291.858&$-$0.675& \object{GAL 291.86-00.68 }&A&84&25&1\\
293.027&$-$1.031& \object{GAL 293.03-01.03 }&D&18&66&1\\
295.760&$-$0.200& \object{GAL 295.76-00.20 }&E&5&17&1\\
296.593&$-$0.975& \object{GAL 296.59-00.98 }&E&8&25&1\\
297.506&$-$0.765& \object{GAL 297.51-00.77 }&B&45&23&1\\
297.655&$-$0.977& \object{GAL 297.66-00.98 }&B&41&26&1\\
298.187&$-$0.782& \object{GAL 298.19-00.78 }&B&14&16&1\\
298.228&$-$0.331& \object{GAL 298.23-00.33 }&A&84&31&1\\
298.868&$-$0.432& \object{GAL 298.87-00.43 }&A&417&25&1\\
299.016&$+$0.148& \object{GAL 299.02+00.15 }&D&15&23&1\\
299.363&$-$0.257& \object{GAL 299.36-00.26 }&D&24&$-$37&1\\
300.479&$-$0.192& \object{GAL 300.48-00.19 }&C&20&26&1\\
300.956&$+$1.161& \object{GUM 43 }&A&60&$-$47&1\\
301.109&$+$0.969& \object{GAL 301.11+00.97 }&A&119&$-$42&1\\
301.814&$+$1.077& \object{GAL 301.81+01.08 }&E&11&$-$42&1\\
302.025&$-$0.044& \object{GAL 302.03-00.04 }&A&19&$-$27&1\\
302.504&$-$0.749& \object{GAL 302.69+00.19 }&D&19&31&1\\
302.690&$+$0.190& \object{WR 47b }&E&13&$-$33&1\\
302.804&$+$1.306& \object{GAL 302.80+01.31 }&D&8&$-$32&1\\
305.173&$-$0.368& \object{GAL 305.17-00.37 }&E&20&$-$45&1\\
305.202&$+$0.022& \object{GAL 305.20+00.02 }&A&241&$-$40&1\\
305.363&$+$0.179& \object{GAL 305.36+00.18 }&A&513&$-$38&1\\
305.537&$+$0.338& \object{GAL 305.54+00.34 }&C&72&$-$39&1\\
305.551&$-$0.005& \object{GAL 305.55-00.01 }&A&117&$-$45&1\\
305.787&$+$0.140& \object{GAL 305.79+00.14 }&E&11&$-$43&1\\
306.256&$+$0.066& \object{GAL 306.26+00.07 }&E&7&$-$37&1\\
306.315&$-$0.361& \object{GAL 306.32-00.36 }&B&40&$-$16&1\\
307.569&$-$0.616& \object{GAL 307.57-00.62 }&D&21&$-$40&1\\
307.620&$-$0.320& \object{GAL 307.62-00.32 }&A&76&$-$37&1\\
308.092&$-$0.432& \object{GAL 308.09-00.43 }&E&21&$-$17&1\\
308.647&$+$0.579& \object{GAL 308.65+00.58 }&A&42&$-$50&1\\
309.057&$+$0.186& \object{GAL 309.06+00.19 }&E&12&$-$47&1\\
309.548&$-$0.737& \object{GAL 309.55-00.74 }&C&35&$-$43&1\\
309.905&$+$0.373& \object{GAL 309.91+00.37 }&B&36&$-$55&1\\
310.176&$-$0.131& \object{GAL 310.18-00.13 }&E&16&4&1\\
310.796&$-$0.408& \object{NAME Kes 20A }&D&17&33,-57&1\\
310.994&$+$0.389& \object{GAL 310.99+00.39 }&E&11&$-$51&1\\
311.114&$-$0.270& \object{GAL 311.11-00.27 }&C&24&36&1\\
311.197&$+$0.752& \object{GAL 311.20+00.75 }&C&19&$-$57&1\\
311.489&$+$0.368& \object{IRAS 14000-6104 }&C&39&$-$59&1\\
311.497&$-$0.483& \object{GAL 311.50-00.48 }&D&27&$-$47&1\\
311.627&$+$0.270& \object{GAL 311.63+00.27 }&B&43&$-$61&1\\
311.852&$-$0.222& \object{GAL 311.85-00.22 }&D&11&$-$55&1\\
311.894&$+$0.100& \object{[CH87] 311.894+0.100 }&A&104&$-$47&1\\
311.922&$+$0.229& \object{GAL 311.92+00.23 }&A&76&$-$45&1\\
312.112&$+$0.314& \object{PN G312.1+00.3 }&A&42&$-$49&1\\
312.953&$-$0.449& \object{GAL 312.95-00.45 }&D&20&$-$47&1\\
313.446&$+$0.176& \object{GAL 313.45+00.18 }&B&26&$-$5&1\\
314.228&$+$0.437& \object{GAL 314.23+00.44 }&A&66&$-$63&1\\
315.312&$-$0.273& \object{GAL 315.31-00.27 }&D&10&16&1\\
316.156&$-$0.492& \object{GAL 316.16-00.49 }&A&35&$-$60&1\\
316.393&$-$0.356& \object{GAL 316.39-00.36 }&C&19&3&1\\
316.808&$-$0.037& \object{GAL 316.81-00.04 }&A&462&$-$36&1\\
317.037&$+$0.300& \object{GAL 317.04+00.30 }&C&40&$-$49&1\\
317.291&$+$0.268& \object{GAL 317.29+00.27 }&D&23&$-$51&1\\
317.598&$-$0.363& \object{GAL 317.60-00.36 }&E&17&$-$38&1\\
317.988&$-$0.759& \object{GAL 317.99-00.76 }&E&13&$-$37&1\\
318.058&$-$0.459& \object{IRAS 14518-5925 }&E&10&$-$31,37&1\\
318.911&$-$0.181& \object{GAL 318.91-00.18 }&A&41&$-$29&1\\
319.157&$-$0.423& \object{GAL 319.16-00.42 }&A&49&$-$22&1\\
319.380&$-$0.025& \object{GAL 319.38-00.03 }&A&91&$-$14&1\\
319.874&$+$0.770& \object{GAL 319.87+00.77 }&A&35&$-$38&1\\
320.109&$-$0.510& \object{[CH87] 320.109-0.510 }&D&16&$-$13&1\\
320.153&$+$0.780& \object{GAL 320.15+00.78 }&A&170&$-$36&1\\
320.236&$+$0.417& \object{GAL 320.24+00.42 }&B&31&$-$31&1\\
320.252&$-$0.332& \object{GAL 320.25-00.33 }&A&92&$-$68&1\\
320.317&$-$0.208& \object{GAL 320.32-00.21 }&A&120&$-$11&1\\
320.379&$+$0.139& \object{GAL 320.38+00.14 }&B&30&$-$3&1\\
320.706&$+$0.197& \object{GAL 320.71+00.20 }&D&20&$-$7&1\\
321.038&$-$0.519& \object{GAL 321.04-00.52 }&A&111&$-$61&1\\
321.105&$-$0.549& \object{GAL 321.11-00.55 }&A&60&$-$56&1\\
321.710&$+$1.157& \object{GAL 321.71+01.16 }&A&38&$-$32&1\\
322.153&$+$0.613& \object{GAL 322.15+00.61 }&A&281&$-$52&1\\
322.407&$+$0.221& \object{GAL 322.41+00.22 }&C&14&$-$30&1\\
324.120&$-$0.954& \object{GAL 324.12-00.95 }&E&6&$-$67&1\\
324.147&$+$0.231& \object{[CH87] 324.147+0.231 }&C&31&$-$91&1\\
324.192&$+$0.109& \object{[CH87] 324.192+0.109 }&A&53&$-$92&1\\
324.954&$-$0.584& \object{[CH87] 324.954-0.584 }&E&10&25&1\\
326.141&$-$0.328& \object{[CH87] 326.141-0.328 }&E&9&$-$65&1\\
326.230&$+$0.976& \object{[CH87] 326.230+0.976 }&E&10&$-$42&1\\
326.441&$+$0.914& \object{[CH87] 326.441-0.396 }&A&161&$-$39&1\\
326.645&$+$0.589& \object{[DBS2003] 95 }&A&425&$-$44&1\\
326.959&$+$0.031& \object{GAL 326.96+00.03 }&C&30&$-$64&1\\
327.313&$-$0.536& \object{[CH87] 327.313-0.536 }&A&722&$-$48&1\\
327.612&$-$0.354& \object{[CH87] 327.612-0.354 }&B&46&$-$72&1\\
327.759&$-$0.351& \object{[CH87] 327.759-0.351 }&A&68&$-$72&1\\
327.834&$+$0.113& \object{[CH87] 327.834+0.113 }&D&18&$-$100&1\\
327.985&$-$0.086& \object{GAL 327.99-00.09 }&A&59&$-$45&1\\
328.310&$+$0.448& \object{GAL 328.31+00.45 }&D&8&$-$97&1\\
328.593&$-$0.518& \object{GAL 328.59-00.52 }&A&208&$-$51&1\\
328.806&$-$0.083& \object{GAL 328.81-00.08 }&B&32&$-$47&1\\
328.812&$+$0.637& \object{GAL 328.81+00.64 }&C&17&$-$42&1\\
329.353&$+$0.144& \object{GAL 329.35+00.14 }&B&34&$-$107&1\\
329.489&$+$0.207& \object{GAL 329.49+00.21 }&D&30&$-$102&1\\
330.041&$-$0.045& \object{GAL 330.04-00.05 }&B&28&$-$38&1\\
330.305&$-$0.385& \object{GAL 330.31-00.39 }&B&19&$-$76&1\\
330.677&$-$0.396& \object{GAL 330.68-00.40 }&A&58&$-$61&1\\
330.856&$-$0.371& \object{GAL 330.86-00.37 }&A&173&$-$56&1\\
331.026&$-$0.152& \object{GAL 331.03-00.15 }&B&59&$-$89&1\\
331.110&$-$0.506& \object{GAL 331.11-00.51 }&A&68&$-$68&1\\
331.259&$-$0.186& \object{GAL 331.26-00.19 }&A&86&$-$85&1\\
331.314&$-$0.336& \object{GAL 331.31-00.34 }&A&125&$-$64&1\\
331.353&$-$0.013& \object{GAL 331.35-00.01 }&A&91&$-$81&1\\
331.354&$+$1.072& \object{GAL 331.35+01.07 }&A&55&$-$79&1\\
331.360&$+$0.507& \object{GAL 331.36+00.51 }&D&19&$-$46&1\\
331.517&$-$0.069& \object{GAL 331.5-00.0 }&A&466&$-$89&1\\
332.148&$-$0.446& \object{GAL 332.15-00.45 }&A&275&$-$55&1\\
332.541&$-$0.111& \object{GAL 332.54-00.11 }&F&10&$-$50&1\\
332.662&$-$0.607& \object{GAL 332.66-00.61 }&A&149&$-$48&1\\
332.978&$+$0.792& \object{GAL 332.98+00.79 }&A&114&$-$52&1\\
333.114&$-$0.441& \object{GAL 333.11-00.44 }&A&303&$-$52&1\\
333.168&$-$0.081& \object{GAL 333.17-00.08 }&A&61&$-$91&1\\
333.292&$-$0.371& \object{GAL 333.29-00.37 }&A&323&$-$50&1\\
333.6&$-$0.1& \object{GAL 333.6-00.1}&A&89&$-$53.7&4\\
333.61&$-$0.208& \object{SNR G333.6-00.2 }&A&840&$-$46&1\\
333.684&$-$0.457& \object{GAL 333.68-00.46 }&C&28&$-$50&1\\
334.529&$+$0.825& \object{GAL 334.53+00.83 }&E&11&$-$77&1\\
334.684&$-$0.107& \object{GAL 334.68-00.11 }&D&24&$-$32&1\\
334.714&$-$0.665& \object{GAL 334.71-00.67 }&A&30&16&1\\
335.748&$-$0.134& \object{GAL 335.75-00.13 }&C&47&$-$52&1\\
335.978&$+$0.185& \object{GAL 335.98+00.19 }&E&24&$-$79&1\\
336.375&$-$0.131& \object{GAL 336.38-00.13 }&B&49&$-$88&1\\
336.404&$-$0.234& \object{GAL 336.40-00.23 }&B&73&$-$93&1\\
336.456&$+$0.038& \object{GAL 336.46+00.04 }&E&22&$-$63&1\\
336.489&$-$0.154& \object{GAL 336.49-00.15}&D&42&$-$84&1\\
336.732&$+$0.072& \object{GAL 336.73+00.07}&B&122&$-$78,-112&1\\
336.840&$+$0.047& \object{GAL 336.84+00.05}&B&109&$-$79&1\\
336.9&$-$0.1& \object{[HHB99] 336.934-0.146}&B&57&$-$73.1&4\\
337.147&$-$0.181& \object{GAL 337.15-00.18 }&A&287&$-$73&1\\
337.3&$-$0.1& \object{[KC97c] G337.3-00.1}&C&39&&4\\
337.548&$-$0.304& \object{GAL 337.55-00.30 }&E&11&$-$101&1\\
337.665&$-$0.048& \object{GAL 337.67-00.05}&C&29&$-$55&1\\
337.949&$-$0.476& \object{GAL 337.95-00.48 }&A&380&$-$40&1\\
338.014&$-$0.121& \object{GAL 338.01-00.12 }&B&46&$-$54&1\\
338.131&$-$0.173& \object{GAL 338.13-00.17 }&A&61&$-$53&1\\
338.398&$+$0.164& \object{GAL 338.40+00.16 }&A&121&$-$29&1\\
338.407&$-$0.238& \object{GAL 338.41-00.24 }&A&98&2&1\\
338.450&$+$0.061& \object{GAL 338.45+00.06 }&A&150&$-$37&1\\
338.742&$+$0.641& \object{GAL 338.74+00.64 }&D&23&$-$62&1\\
338.921&$-$0.089& \object{GAL 338.92-00.09 }&A&60&$-$40&1\\
338.943&$+$0.604& \object{GAL 338.94+00.60 }&A&87&$-$63&1\\
339.089&$-$0.216& \object{GAL 339.09-00.22 }&D&18&$-$120&1\\
339.128&$-$0.408& \object{SNR G339.2-00.4 }&D&24&$-$37&1\\
339.286&$+$0.163& \object{GAL 339.29+00.16 }&E&14&$-$71&1\\
339.578&$-$0.124& \object{GAL 339.58-00.12 }&D&21&$-$30&1\\
339.838&$+$0.274& \object{GAL 339.84+00.27 }&A&35&$-$19&1\\
339.955&$-$0.566& \object{GAL 339.96-00.57 }&E&10&$-$89&1\\
340.047&$-$0.253& \object{GAL 340.05-00.25 }&C&34&$-$52&1\\
340.240&$+$0.482& \object{GAL 340.24+00.48 }&C&13&$-$62&1\\
340.279&$-$0.222& \object{GAL 340.28-00.22 }&A&77&$-$43&1\\
340.777&$-$1.008& \object{GAL 340.78-01.01 }&A&233&$-$25&1\\
341.050&$-$0.100& \object{GAL 341.05-00.10 }&E&6&$-$38&1\\
341.264&$-$0.317& \object{GAL 341.26-00.32}&E&24&$-$38&1\\
341.963&$+$0.205& \object{GAL 341.96+00.21 }&D&12&$-$8&1\\
342.085&$+$0.423& \object{GAL 342.09+00.42 }&A&85&$-$65&1\\
342.300&$+$0.314& \object{GAL 342.30+00.31 }&E&25&$-$122&1\\
342.382&$-$0.044& \object{GAL 342.38-00.04 }&D&26&$-$13&1\\
345.215&$-$0.749& \object{GAL 345.22-00.75 }&D&15&$-$18&1\\
345.231&$+$1.035& \object{GAL 345.23+01.04 }&A&211&$-$9&1\\
345.404&$+$1.406& \object{GAL 345.40+01.41 }&A&27&$-$15&1\\
345.450&$+$0.209& \object{GAL 345.45+00.21 }&B&32&$-$13&1\\
345.495&$+$0.326& \object{GAL 345.50+00.33 }&A&38&$-$20&1\\
345.555&$-$0.042& \object{GAL 345.56-00.04 }&A&76&$-$6&1\\
345.645&$+$0.010& \object{GAL 345.65+00.01 }&A&81&$-$10&1\\
345.722&$+$0.153& \object{HRDS G345.722+0.153 }&E&6&$-$77.9&2\\
345.827&$+$0.041& \object{GAL 345.83+00.04 }&B&23&$-$10&1\\
346.056&$-$0.021& \object{HRDS G346.056-0.021 }&D&9&$-$76.8, -3.4&2\\
346.077&$-$0.056& \object{HRDS G346.077-0.056 }&E&12&$-$84.7,-7.9&2\\
346.206&$-$0.071& \object{GAL 346.21-00.07 }&C&18&$-$108&1\\
346.267&$+$0.128& \object{HRDS G346.267+0.128 }&E&5&$-$32.2&2\\
346.530&$-$0.013& \object{HRDS G346.530-0.013 }&E&5&0.6&2\\
346.539&$+$0.097& \object{GAL 346.54+00.10 }&B&35&2&1\\
346.875&$+$0.328& \object{HRDS G346.875+0.328 }&E&10&4.1&2\\
347.386&$+$0.266& \object{GAL 347.39+00.27 }&E&20&$-$97&1\\
347.600&$+$0.211& \object{GAL 347.60+00.21 }&E&19&$-$96&1\\
347.772&$+$0.131& \object{HRDS G347.772+0.131 }&E&6&$-$88.7&2\\
347.893&$+$0.044& \object{GAL 347.89+00.04 }&A&51&$-$31&1\\
347.918&$-$0.761& \object{HRDS G347.918-0.761 }&E&6&6.1&2\\
347.964&$-$0.439& \object{GAL 347.96-00.44 }&C&27&$-$97&1\\
348.061&$+$0.242& \object{HRDS G348.061+0.242 }&E&5&0.7&2\\
348.148&$+$0.255& \object{HRDS G348.148+0.255 }&E&5&$-$66.3, -1.4&2\\
348.225&$+$0.459& \object{GAL 348.23+00.46 }&B&61&$-$12&1\\
348.231&$-$0.982& \object{ESO 333-3 }&A&237&$-$18&1\\
348.557&$-$0.985& \object{HRDS G348.557-0.985 }&E&17&$-$10.5&2\\
348.715&$-$1.031& \object{GAL 348.72-01.03 }&A&562&$-$13&1\\
348.891&$-$0.179& \object{HRDS G348.891-0.179 }&E&8&10.1&2\\
349.111&$+$0.105& \object{GAL 349.11+00.11 }&C&33&$-$74&1\\
349.140&$+$0.020& \object{GAL 349.14+00.02 }&B&81&$-$92, 17&1\\
349.216&$+$0.144& \object{HRDS G349.216+0.144 }&D&11&$-$65.7&2\\
349.579&$-$0.68& \object{HRDS G349.579-0.680 }&C&12&$-$19.4&2\\
349.84&$-$0.537& \object{GAL 349.84-00.54 }&A&110&$-$25&1\\
350.004&$+$0.438& \object{HRDS G350.004+0.438 }&E&13&$-$33.5&2\\
350.129&$+$0.088& \object{GAL 350.13+00.09 }&A&143&$-$69&1\\
350.177&$+$0.017& \object{HRDS G350.177+0.017 }&D&14&$-$68.9&2\\
350.33&$+$0.157& \object{HRDS G350.330+0.157 }&E&11&$-$62.9&2\\
350.335&$+$0.107& \object{[KC97c] G350.3+00.1}&E&12&$-$66.1&3\\
350.524&$+$0.960& \object{GAL 350.52+00.96 }&A&77&$-$10&1\\
350.769&$-$0.075& \object{HRDS G350.769-0.075 }&E&5&$-$62.6&2\\
350.813&$-$0.019& \object{GAL 350.81-00.02 }&A&29&$-$5&1\\
350.996&$-$0.577& \object{GAL 351.00-00.58 }&E&11&$-$17&1\\
351.063&$+$0.662& \object{[L89b] 351.063+00.662}&A&150&$-$3.8&3\\
351.192&$+$0.708& \object{[L89b] 351.192+00.708}&A&224&$-$3.4&3\\
351.201&$+$0.483& \object{[L89b] 351.201+00.483}&A&157&$-$7.1&3\\
351.358&$+$0.666& \object{GAL 351.36+00.67 }&A&566&$-$3&1\\
351.359&$+$1.014& \object{HRDS G351.359+1.014 }&E&11&$-$8.6&2\\
351.467&$-$0.462& \object{GAL 351.47-00.46 }&A&80&$-$21&1\\
351.601&$-$0.348& \object{GAL 351.60-00.35 }&C&20&$-$94&1\\
351.617&$+$0.171& \object{GAL 351.62+00.17 }&A&173&$-$43&1\\
351.641&$-$1.256& \object{GAL 351.60-01.30 }&A&102&$-$13&1\\
351.662&$+$0.518& \object{GAL 351.66+00.52 }&D&19&$-$2&1\\
351.691&$+$0.669& \object{HRDS G351.691+0.669 }&B&25&3.1&2\\
351.694&$-$1.165& \object{GAL 351.69-01.17 }&A&138&$-$13&1\\
352.398&$-$0.057& \object{GAL 352.40-00.06 }&D&21&$-$89&1\\
352.521&$-$0.144& \object{HRDS G352.521-0.144 }&B&61&$-$57.3,-38&2\\
352.61&$+$0.177& \object{HRDS G352.610+0.177 }&E&13&$-$50.4&2\\
352.611&$-$0.172& \object{GAL 352.61-00.17 }&B&21&$-$82&1\\
352.676&$+$0.148& \object{GAL 352.68+00.15 }&E&17&$-$45&1\\
353.035&$+$0.748& \object{[L89b] 353.035+00.748}&C&44&$-$9.1&3\\
353.136&$+$0.660& \object{GAL 353.14+00.66 }&A&162&$-$4&1\\
353.186&$+$0.887& \object{[L89b] 353.186+00.887}&A&95&$-$4.7&3\\

\enddata
\tablecomments{Columns are as follows: 1) Galactic longitude, 2) Galactic latitude, 3) SIMBAD identification, 4) quality factor, 5) $T_{\rm{cont}}$ (K), 6) RRL velocity, 7) reference for \HII region coordinates and velocity.}
\tablerefs{1: \citet{CH87}, 2: \citet{GBTHRDS}, 3: \citet{L89}, 4: \citet{Wilson70}}
\end{deluxetable}
\clearpage

\begin{figure}[h]
\centering
\includegraphics[width=\columnwidth]{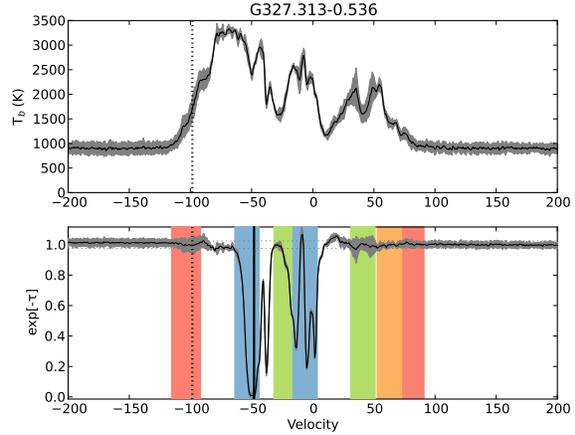}
\caption{\HI emission/absorption spectrum pairs. In each figure, the top panel shows the emission spectra. The emission is shown by the solid line (this is the average of the three “off” positions, see Section \ref{extract}) and the emission envelope ($3\sigma_{T_{\rm{off}}}$) is shown in grey. Absorption, $e^{-\tau}$, is displayed in the bottom panel. The \HI absorption spectrum  (see Equation \ref{EqAbs}) is shown by the solid line and the grey envelope signifies $\Delta_{\textrm{abs}}$ (calculated from the emission envelope). The absorption panel also displays the fluctuation in the baseline of the absorption spectrum ($\sigma_{\rm{rms}}$) (horizontal dotted lines). The \HII region name and reference are shown as well as the expected velocity ranges of Galactic structure features with the same color system as Figure \ref{Fig5}. (A color version and the complete figure set (252 images) of this figure are available in the online journal.) \label{Fig4}}
\end{figure}

\paragraph{Spectrum Figures and Data}
For every included \HII region from Table \ref{Tablecatalog}, the emission and absorption spectra (along with associated uncertainties) are displayed in Figure \ref{Fig4}.  In each Figure, the top panel shows the emission spectra.  The emission is shown by the solid line ($T_{\rm{off}}$, this is the average of the three `off' positions, see \S \ref{extract}) and the emission envelope ($3\sigma_{T_{\rm{off}}}$: standard deviations between the `off' positions) is shown in grey.  Absorption, $e^{-\tau}$, is displayed in the bottom panel.  The \HI absorption spectrum (see Equation \ref{EqAbs}) is shown by the solid line and the grey envelope signifies $\Delta_{\textrm{abs}}$ (calculated from the emission envelope). The absorption panel also displays the fluctuation in the baseline of the absorption spectrum ($\sigma_{rms}$) (horizontal dotted lines).

The \HII region name and reference are shown as well as the expected velocity ranges of spiral arm features with the same `crayon' color system as Figure \ref{Fig5}.  This color coding provides an accessible method of `reading\rq{} the \HI absorption spectra in terms of known Galactic features.\\

The \HI emission/absorption spectra, for each included \HII region, are also available in a tar file.  We provide an example of this data in Table \ref{TableSpectrumEG}.  For each velocity channel we provide the on source brightness temperature ($T_{\rm{on}}$); average of the three off source brightness temperatures (i.e. the emission spectrum, $T_{\rm{off}}$);  the uncertainty in $T_{\rm{off}}$, caused by the differences in the three off source positions,$3\sigma_{T_{\rm{off}}}$ ; the absorption value ($e^{-\tau}$); and the $\Delta_{\textrm{abs}}$ absorption uncertainty envelope.

\begin{table}[h!]
\centering
\begin{tabular}{|cccccc|}
\hline
[1]&[2]&[3]&[4]&[5]&[6]\\
\hline
-10.718	&	5.042	&	3.638	&	0.803	&	1.008	&	0.079	\\
-9.894	&	4.320	&	3.266	&	1.192	&	1.006	&	0.067	\\
-9.069	&	4.932	&	4.621	&	0.758	&	1.002	&	0.078	\\
-8.245	&	6.187	&	4.059	&	0.850	&	1.012	&	0.099	\\
-7.420	&	6.265	&	6.871	&	1.706	&	0.997	&	0.100	\\
-6.596	&	6.783	&	7.486	&	0.352	&	0.996	&	0.109	\\
-5.771	&	9.314	&	7.669	&	0.612	&	1.009	&	0.152	\\
-4.947	&	8.126	&	8.502	&	1.465	&	0.998	&	0.132	\\
-4.122	&	9.135	&	8.663	&	0.611	&	1.003	&	0.149	\\
-3.298	&	11.130	&	11.743	&	0.890	&	0.997	&	0.183	\\
-2.473	&	10.098	&	14.215	&	1.997	&	0.977	&	0.165	\\
-1.649	&	10.418	&	15.320	&	0.644	&	0.972	&	0.171	\\
-0.824	&	12.797	&	16.794	&	1.634	&	0.977	&	0.211	\\
0.000	&	15.127	&	20.274	&	2.506	&	0.971	&	0.250	\\
0.825	&	18.053	&	23.362	&	2.692	&	0.970	&	0.300	\\
1.649	&	15.809	&	28.417	&	1.363	&	0.929	&	0.262	\\
2.474	&	6.041	&	33.051	&	3.283	&	0.848	&	0.096	\\
3.298	&	-2.430	&	39.661	&	6.883	&	0.762	&	0.047	\\
4.123	&	17.582	&	42.043	&	5.126	&	0.862	&	0.292	\\
4.947	&	28.948	&	44.790	&	3.800	&	0.911	&	0.484	\\
5.772	&	33.344	&	44.136	&	2.939	&	0.939	&	0.559	\\
6.596	&	36.165	&	44.364	&	2.387	&	0.954	&	0.607	\\
7.421	&	28.446	&	45.145	&	3.021	&	0.906	&	0.476	\\
8.245	&	10.029	&	43.982	&	4.989	&	0.808	&	0.164	\\
9.069	&	15.144	&	46.651	&	3.893	&	0.822	&	0.251	\\
9.894	&	30.727	&	48.985	&	3.917	&	0.897	&	0.515	\\
10.718	&	37.734	&	48.884	&	3.879	&	0.937	&	0.633	\\

\hline
\end{tabular}
\caption{Example of spectrum data.  Columns are as follows: [1] velocity channel (\kms); [2] $T_on$ (K); [3] average $T_{off}$ (K); [4] uncertainty in $T_{off}$, $3\sigma_{T_{\rm{off}}}$ (K); [5] absorption, $e^{-\tau}$; [6] uncertainty in absorption, $\Delta_{\textrm{abs}}$.  This Table, and similar files for each \HII region included in the catalog, are published in entirety in the electronic edition of the journal, a portion is shown here for guidance regarding the form and content of each file.}
\label{TableSpectrumEG}
\end{table}

\section{Discussion}
\label{Discussion}
Here we briefly discuss the global properties of the \HI absorption catalog (\S \ref{disscusslv}), as well as an example of its use - investigating the $lv$ locus of the Near 3kpc Arm (see \S \ref{N3kpc}).

\subsection{Longitude-Velocity Distribution}\label{disscusslv}
\begin{figure*}[t]
\centering
\includegraphics[width=\textwidth]{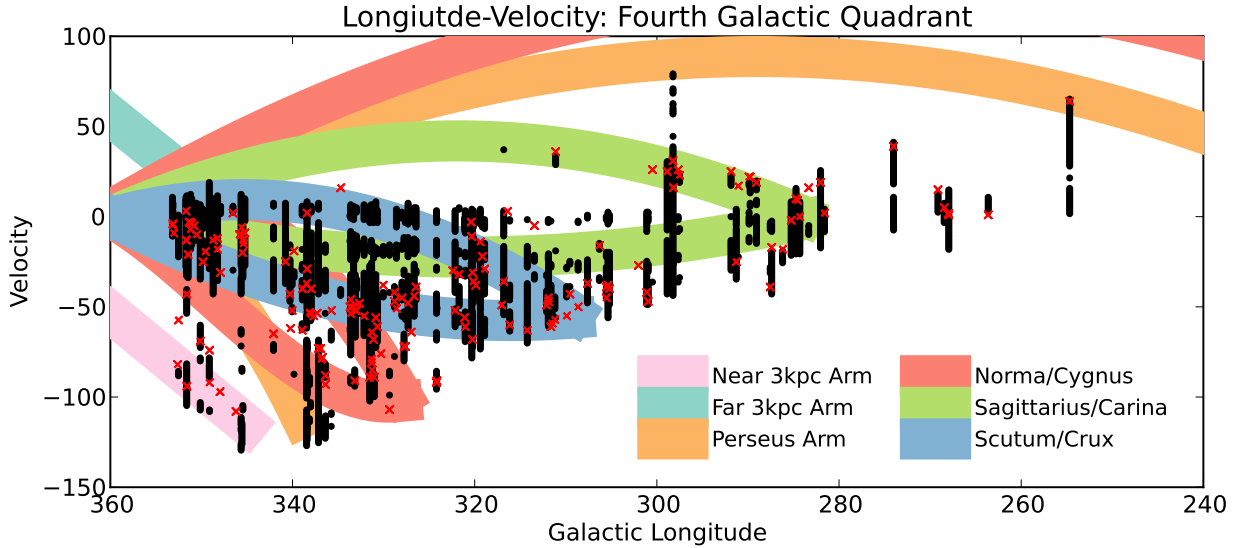}
\caption{\HII region RRL velocities (red crosses) and \HI absorption (black dots) overlaid on a longitude-velocity `crayon diagram'.  Each dot denotes the central velocity of five adjacent velocity channels ($\sim5$\kms), with each channel demonstrating a statistically significant amount of \HI absorption.  Only \HII regions with \HI absorption spectra of Quality 'C' or better are included for use in this figure. (A color version of this figure is available in the online journal.)}\label{Fig5}
\end{figure*}

One simple application of this catalog is an investigation of the distribution of \HI absorption in longitude-velocity ($lv$) space.  Firstly, we construct an `$lv$ crayon diagram\rq{} (Figure \ref{Fig5}) of known Galactic structures.  The Near and Far 3kpc Arm fits are provided by \citet{DameThaddeus08} and the fourth quadrant spiral arms are taken from \citet{Vallee08} (who represents the Distant Arm of \citet{McCGDistantArm} as the Cygnus Arm beyond the Solar Circle).  The velocity width of each crayon feature is set to 20\kms.

For each source in the catalog (of quality factor C or better), we plot both the systemic velocity of the \HII region and any associated \HI absorption (Figure \ref{Fig5}).  It is clear that both the \HII regions and their associated \HI absorption trace the spiral arm structures, especially in the inner Galaxy.  This is in keeping with \citet{Jones13} who find that $>90\%$ of their \HII region sample is associated with known spiral arm structures.  \HII regions are considered to be the archetypical tracers of Galactic spiral structure \citep{GBTHRDS}; but cold, dense gas, traced by \HI absorption is also more likely to be located within the spiral arms than in the inter-arm region.  Each black marker on the Figure corresponds to five consecutive velocity channels which display significant absorption.  It is clear that the \HI absorption is also associated with the spiral arms.  This is to be expected, as dense, cold gas which is traced by \HI absorption is more likely to be located within the spiral arms than in the inter-arm regions.  We find \HI absorption associated with all inner Galaxy spiral arms.

Figure \ref{Fig5} not only demonstrates the suitability of both data-sets (\HII region velocities and \HI absorption) as spiral arm tracers, but also their inherent velocity uncertainty distributions.  Galactic streaming motions are often estimated to be on the order of $\lesssim15$\kms \citep{BaniaLockman84, Kolpak03}, i.e. within the velocity width of the crayon diagram features.  However, \HI absorption associated with \HII regions can extend $\sim20$\kms beyond the systemic velocity of the region \citep{Dickey03, Jones13}.  

The association of \HI absorption with a Galactic structure feature can be used to infer lower distance limits, or the location of the \HII region in our Galaxy \citep{JonesDickey12, Jones13}.

\subsubsection{\HI Absorption in the Outer Galaxy}
From Figure \ref{Fig5} it is evident that this catalog is primarily limited to the inner Galaxy (within the Solar circle); unlike, for example, the work of \citet{Strasser07}.  

\citet{Strasser07} used the SGPS and the E/A spectrum extraction method to study \HI absorption towards 111 extragalactic continuum sources - in order to investigate absorption from the outermost arms of the Milky Way.  Of these 111 sources, only 17 demonstrated any signature of \HI absorption at positive velocities; corresponding to a location in the outer Galaxy.  

Because the \citet{Strasser07} source sample included strictly extra-Galactic sources, the \HI absorption spectra reflect lines of sight through the entire Galactic plane.  Whereas spectra from this study reflect lines of sight towards the \HII regions themselves, located within the Galaxy.  For this reason, this work is far less likely to identify \HI absorption in the outermost spiral arms in the fourth quadrant - as the number of \HII regions known (with RRL velocities) is extremely limited in the most distant features.  

Nevertheless, for longitudes $l<300\arcdeg$ the line of sight distance to the Solar circle is relatively small and investigations of \HI absorption in the outer Galaxy become possible; although at these longitudes, there are very few known \HII regions (see Figure \ref{Fig5}).  Regions with velocities $\lesssim15$\kms correspond to very short line of sight distances.  At $l\approx 250\arcdeg$ the near-continuous absorption profile is associated with lines of sight through the \objectname[LOCAL ARM]{Local Arm}.

\subsection{Locus of the Near 3kpc Arm \label{N3kpc}}

\begin{figure}[h!]
\centering
\includegraphics[width=\columnwidth]{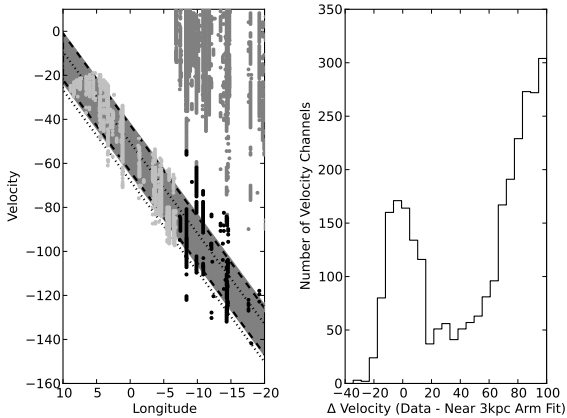}
\caption{\HI absorption associated with the Near 3kpc Arm.  \textit{Left Panel}: $lv$ diagram of \HI absorption corresponding to the Near 3kpc Arm.  Light grey dots are data from \citet{Jones13}, dark grey dots are from this work (but were not included in the linear fit analysis), black dots are also from this work (and were used in the analysis of the locus of the Arm). \textit{Right Panel:} fit of data to model (Equation \ref{Eqboth}).  The histogram displays the difference in velocity between Equation \ref{Eqboth} and each channel that displays significant absorption (see Figure \ref{Fig5}).  Note only regions with $l>340\arcdeg$ are considered here.\label{Fig6}}
\end{figure}
The investigation of the $lv$ distribution of \HI absorption (see Figure \ref{Fig5}) can be further specialized to individual features.  In \citet{Jones13} the locus of the Near and Far 3kpc Arms (in $lv$ space) was investigated using \HI absorption.  This work presents an opportunity to extend this analysis, by including \HI absorption associated with the Near 3kpc Arm at longitudes $l<350\arcdeg$.  This is a complementary investigation to that first performed by \citet{DameThaddeus08} using CO 1-0 emission.  

Using \HI absorption channels from the Near 3kpc Arm analysis of \citet{Jones13} in addition to those from this work---such that the entire longitude extent 10\arcdeg$>l>340$\arcdeg of the Near 3kpc Arm is included---a linear fit to the $lv$ locus of the Near 3kpc Arm was performed:
\begin{equation}
V=-53.3+4.1l\qquad\Delta V=25.9\textrm{\kms} \label{Eqboth}
\end{equation}
Here $\Delta V$ refers to the FWHM of a Gaussian fit to the velocity profile of the Arm model (see right hand panel of Figure \ref{Fig6}).

With the inclusion of \HI absorption from the longitude range $l<340\arcdeg$, the model for the locus of the Near 3kpc Arm (Equation \ref{Eqboth}) is extremely consistent with the fit provided by \citet{DameThaddeus08} from CO observations; $R^2>0.97$ and the standard error of the mean difference is 5.4\kms, within the FWHM of either fit.

 Figure \ref{Fig6} demonstrates the location of \HI absorption associated with the Arm in $lv$ space, as well as an evaluation of the fit.  The left hand panel,  supplementary to Figure \ref{Fig5}, is an $lv$ diagram extended to $l=10\arcdeg$ over the velocity range of the Near 3kpc Arm.  Channels of significant absorption are shown,  in addition to those from \citet{Jones13}.  The $lv$ linear fits to the locus of the Near 3kpc Arm from \citet[][$V=-53.1+4.6, \Delta V=19.7$ \kms]{DameThaddeus08}, \citet[][$V=-59.2+4.12, \Delta V=17.3$ \kms]{Jones13} and this work (see Equation \ref{Eqboth}) are also shown.  In the right hand panel an evaluation of the fit is made.  All velocity channels shown in the left-hand panel are compared to the linear Near 3kpc Arm fit  (similar to the CO 1-0, Figure 2 of \citet{DameThaddeus08}).  The Near 3kpc Arm is identified as a well defined peak in \HI absorption, centered around the fit to the linear model (however, the scale height of the component is exaggerated by the removal of non-Near 3kpc Arm absorption for $360\arcdeg>l>353\arcdeg$).  The hump at $\Delta V=30$ (i.e. velocities consistent with a line parallel to Equation \ref{Eqboth} $+30$\kms) may be the signature of the Norma Arm, which is approximately parallel to the Near 3kpc Arm in $lv$ space for longitudes $l>340\arcdeg$ (see Figure \ref{Fig5}).  The broad central peak is the result of the remaining disk rotation, both foreground and background to the Near 3kpc Arm.

Note that longitudes $l<350\arcdeg$ the Far 3kpc Arm is confused in velocity with the other inner Galaxy spiral arms - therefore investigation of the locus of the Far 3kpc Arm, extending the work of \citet{Jones13}, is not possible here.

\section{Summary}
The first attempt to test and interpret \HI emission/absorption spectrum pairs from the SGPS (test region) was performed by \citet{Dickey03}.  However, until now, no complete census of \HI absorption towards Galactic continuum sources in the SGPS has been completed.  This paper presents, in graphical, numerical and summary formats, the \HI emission and absorption spectrum pairs from every known Galactic \HII region distinctly detectable in the SGPS (255\arcdeg$<l<$353\arcdeg, $|b|\lesssim1$\arcdeg)---a total of 252 regions.

We have demonstrated one use of this catalog by examining the $lv$ distribution of \HI absorption in the Milky Way---including a re-evaluation of the locus of the Near 3kpc Arm in $lv$ space.  This catalog has the potential to be an integral data set for numerous works; perhaps even sparking the invigoration of the search for \HI spiral arms of the Milky Way in absorption, rather than emission, as \citet{Lockman02} predicted.

\acknowledgements
This research has made use of of NASA's Astrophysics Data System; the SIMBAD database and VizieR catalogue access tool, CDS, Strasbourg, France; and matplotlib for python \citep{matplotlib}. J. R. Dawson is a University Associate of the University of Tasmania.

\end{document}